\documentclass[amsmath,amssymb,superscriptaddress, twocolumn,notitlepage,nofootinbib, aps, prl]{revtex4-2}
\usepackage{bm}
\usepackage[normalem]{ulem}
\usepackage{graphicx}
\usepackage[compact]{titlesec}
\usepackage{xr}
\usepackage{mathtools}
\titlespacing{\section}{4pt}{*4}{*4}
\titlespacing{\subsection}{2pt}{*2}{*2}
\titlespacing{\subsubsection}{2pt}{*2}{*2}
\usepackage{float}
\usepackage{color,soul}
\usepackage[table]{xcolor}
\usepackage[colorlinks=true, linkcolor=blue, citecolor=blue, urlcolor=blue, breaklinks=true]{hyperref}
\usepackage{verbatim} 

\usepackage{microtype}
\usepackage{mathpazo}
\usepackage[semibold]{sourcesanspro}
\usepackage{amsmath}
\usepackage{amssymb}
\usepackage[mathlines]{lineno}
\newcommand{\ket}[1]{|#1\rangle}

\newcommand{\ketbra}[2]{\vert#1\rangle\langle#2\vert}

\newcommand{\blk}{\color{black}}

\renewcommand{\figurename}{\textbf{Fig.}}
\renewcommand{\thefigure}{{\bf \arabic{figure}}}
\renewcommand{\tablename}{\textbf{Tab.}}
\renewcommand{\thetable}{{\bf \arabic{table}}}

\date{\today}

\begin{document} 
\title{A heralded quantum amplifier of multi-photon states \\
}
    
\author{Luis Villegas-Aguilar}\email{luis@villegasaguilar.com}
\author{Farzad Ghafari}
\affiliation{Queensland Quantum and Advanced Technologies Institute and Centre for Quantum Computation and Communication Technology, Griffith University, Yuggera Country, Brisbane, QLD 4111, Australia}

\author{Matthew S. Winnel}
\affiliation{Centre for Quantum Computation and Communication Technology, School of Mathematics and Physics, University of Queensland,
	Brisbane, QLD 4072, Australia}

\author{Varun B. Verma}
\author{Lynden K. Shalm}
\affiliation{National Institute of Standards and Technology, 325 Broadway, Boulder, Colorado 80305, USA}

\author{Timothy C. Ralph}
\affiliation{Centre for Quantum Computation and Communication Technology, School of Mathematics and Physics, University of Queensland,
	Brisbane, QLD 4072, Australia}

\author{Geoff J. Pryde}
\author{Sergei Slussarenko}
\email{s.slussarenko@griffith.edu.au}

\affiliation{Queensland Quantum and Advanced Technologies Institute and Centre for Quantum Computation and Communication Technology, Griffith University, Yuggera Country, Brisbane, QLD 4111, Australia}
\begin{abstract}
	Large-scale quantum networking systems will inevitably require methods to overcome photon loss.
	While the no-cloning theorem forbids perfect and deterministic amplification of unknown quantum states~\cite{caves1981QuantummechanicalNoiseInterferometer, wootters1982SingleQuantumCannot}, probabilistic heralded amplification schemes offer a viable path forward~\cite{ralph2009NondeterministicNoiselessLinear, xiang2010HeraldedNoiselessLineara}.
	Yet, for over a decade, successful multi-photon state amplification has remained out of reach, despite the fundamental importance of such states in achieving quantum advantage in optical applications.
	Here, we experimentally demonstrate a high-fidelity and post-selection-free amplifier for multi-photon states.
	We achieve heralded amplification of states with up to two photons in a single optical mode, with over a hundredfold intensity gain, and verify the coherence-preserving operation of our scheme.
	Our approach is scalable to higher photon numbers and enables noiseless amplification of complex multi-photon quantum states, with applications in large-scale quantum communication systems, distributed quantum metrology, and information processing.
\end{abstract}

\maketitle

Heralded amplification (HA) holds immense promise for quantum technologies, with applications including channel error correction~\cite{ralph2011QuantumErrorCorrection, dias2018QuantumErrorCorrectiona, slussarenko2022QuantumChannelCorrection}, device-independent~\cite{gisin2010ProposalImplementingDeviceIndependenta} and continuous-variable quantum key distribution~\cite{blandino2012ImprovingMaximumTransmission, notarnicola2023LongdistanceContinuousvariableQuantum}, quantum-enhanced metrology~\cite{usuga2010NoisepoweredProbabilisticConcentrationa, zhao2017QuantumEnhancementSignalnoise, xia2019RepeaterenhancedDistributedQuantum}, and quantum repeaters~\cite{dias2017QuantumRepeatersUsing, ghalaii2020LongDistanceContinuousVariableQuantum, seshadreesan2020ContinuousvariableQuantumRepeatera, dias2022DistributingEntanglementFirst,zhao2023EnhancingQuantumTeleportation}.
Its ability to distil and purify entanglement degraded by loss~\cite{xiang2010HeraldedNoiselessLineara} makes it an indispensable tool for scalable quantum communication.
However, current HA demonstrations~\cite{fiurasek2009EngineeringQuantumOperations, zavatta2011HighfidelityNoiselessAmplifier, osorio2012nla,ulanov2015UndoingEffectLoss,bruno2016amplification}, including the original heralded amplifier~\cite{ralph2009NondeterministicNoiselessLinear, xiang2010HeraldedNoiselessLineara} based on the one-photon quantum scissors~\cite{pegg1998OpticalStateTruncation}, are limited to amplifying optical fields with negligible contributions beyond the vacuum and single-photon terms.
This operational constraint means that current HA schemes cannot address the ultimate frontier in optical quantum technologies: scalable, highly-interconnected protocols that rely on high-photon-number states.
For instance, large squeezed states---such as amplitude squeezed cat states and Gottesman-Kitaev-Preskill states~\cite{winnel2024DeterministicPreparationOptical, konno2024LogicalStatesFaulttolerant, aghaeerad2025ScalingNetworkingModular}---are critical for achieving fault-tolerance in continuous-variable optical quantum computing.
Similarly, entangled multi-photon states enable supersensitive quantum metrology beyond the shot-noise limit~\cite{slussarenko2017UnconditionalViolationShot, thekkadath2020QuantumenhancedInterferometryLarge, ferdous2024EmergenceMultiphotonQuantuma, deng2024QuantumenhancedMetrologyLarge} and play a key role in emerging information processing protocols
~\cite{stobinska2019QuantumInterferenceEnables}.
All these schemes will demand mechanisms for multi-photon amplification, especially within the context of networked~\cite{aghaeerad2025ScalingNetworkingModular,alexander2025ManufacturablePlatformPhotonic}, and distributed and long-distance applications~\cite{guo2020DistributedQuantumSensing, main2025DistributedQuantumComputing}.
Consequently, extending the working principles of HA to states with large photon numbers has been a longstanding goal in the field.

In theory, a multi-photon amplifier could be built with several one-photon scissors operating in parallel~\cite{ralph2009NondeterministicNoiselessLinear}.
In practice, one would need simultaneous and coherent operation of many quantum scissors to achieve ideal amplification, ultimately collapsing the protocol's overall success probability.
Other approaches either yield a distorted output state~\cite{jeffers2010NondeterministicAmplifierTwophoton, park2016EfficientNoiselessLinear, fadrny2024ExperimentalPreparationMultiphotonadded}, rely on post-selection rather than heralding~\cite{chrzanowski2014MeasurementbasedNoiselessLinear, zhao2017CharacterizationMeasurementbasedNoiseless}, or are hindered by significant losses in commercially available quantum hardware~\cite{goldberg2023TeleamplificationBorealisBosonsampling}.
As a result, none of these protocols have succeeded in realizing a physical device capable of coherently amplifying quantum states of more than a single photon.

Here, we overcome these challenges and experimentally demonstrate a scheme for ideal multi-photon amplification based on the generalized $n$-photon quantum scissor architecture~\cite{winnel2020GeneralizedQuantumScissors, fiurasek2022OptimalLinearopticalNoiseless, guanzon2022IdealQuantumTeleamplification}.
The main element in our amplifier is a low-loss, multi-port quantum interferometer that can be precisely tuned via the polarization mode of single photons.
We experimentally implement the simplest instance of the device for $n=2$. We demonstrate its coherence-preserving operation for various levels of input signal attenuation and provide both theoretical and experimental analyses of the achievable gain and its sensitivity to optical loss within the circuit. Additionally, we discuss the use of multi-photon heralded amplification in entanglement distillation and quantum state purification protocols, highlighting the post-selection-free amplification of multi-photon states as a vital tool for the next generation of optical quantum technologies.

\begin{figure}[ht]
	\centering
	\includegraphics[width=89mm]{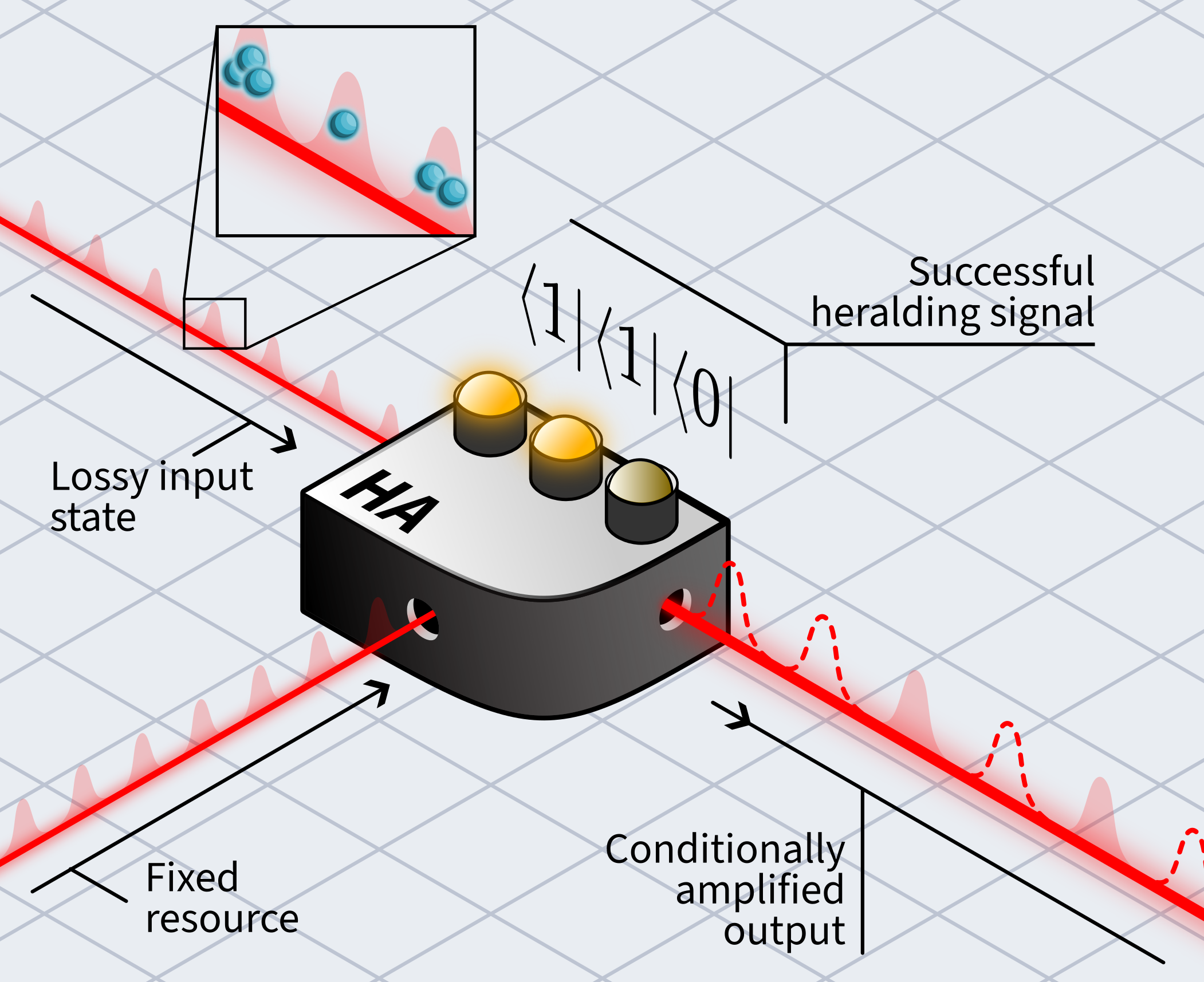}
	\centering
	\caption[]{\textbf{A device for amplifying multi-photon states.}
		An arbitrary quantum optical field (with a finite photon number) is coherently mixed in a multi-mode interferometer with a resource state of $n$ photons.
		Conditioned on detecting $n$ photons in particular detectors at the output, amplification occurs up to the $n$-th order.
		If the input contains more than $n$ photons, the amplifier also truncates it in the photon-number basis to $n$. Otherwise, the input state is perfectly amplified.
	}
	\label{fig:conceptual}
\end{figure}

\begin{figure*}
	\centering
	\includegraphics[width=183mm]{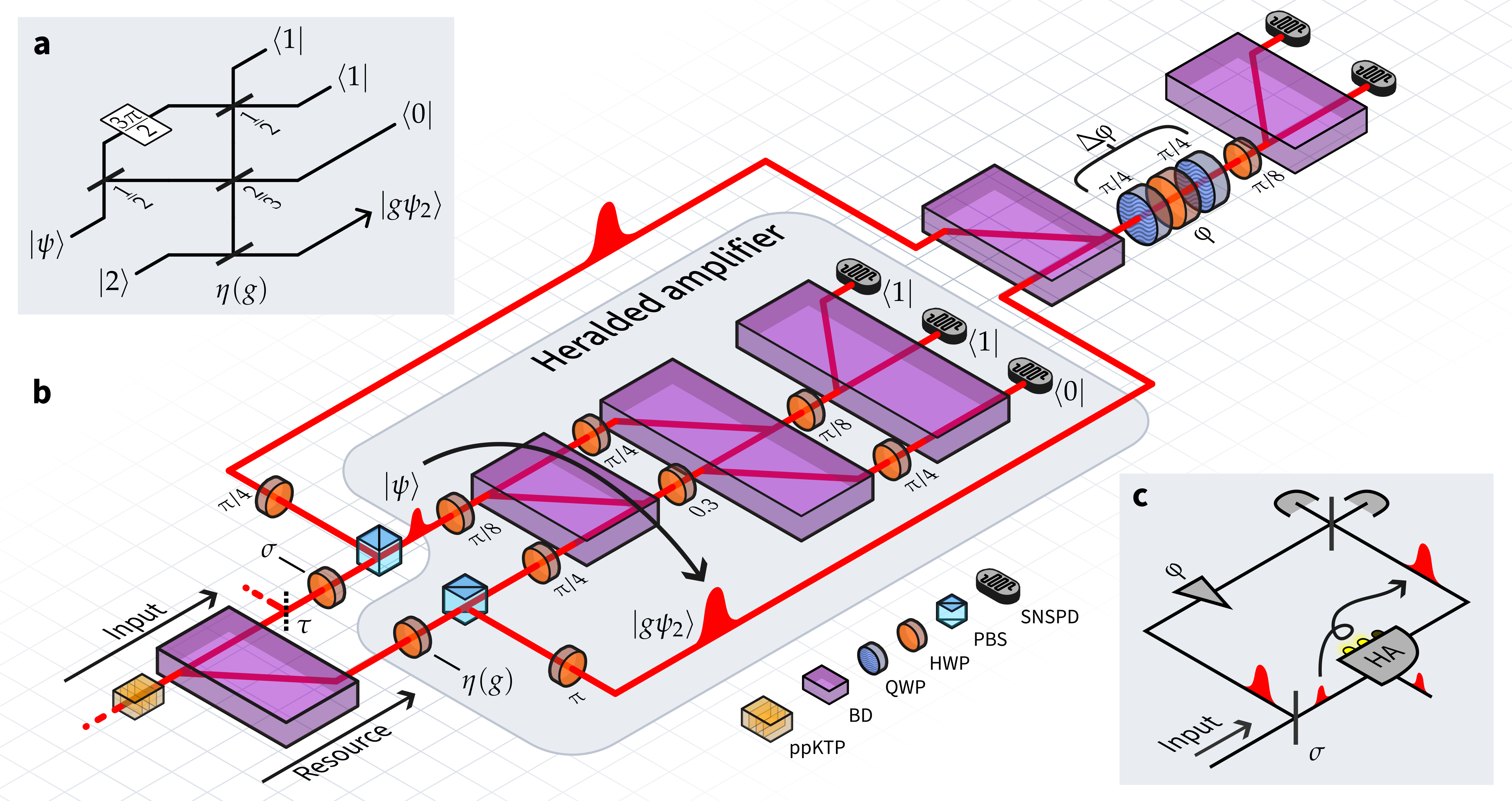}
	\centering
	\caption[]{\textbf{Experimental heralded amplification of two-photon states.}
		\textbf{a}, The quantum circuit for an $n=2$ photon amplifier, featuring a three-mode linear optical circuit composed of beam splitters and a phase shifter. \textbf{b}, A nonlinear crystal generates the input state $\ket{\psi}$ (with negligible $n>3$ terms) and the $\ket{2}$ resource that powers the amplifier.
		The dotted beam splitter ($\tau$) can be used to introduce additional variable loss to the input.
		After splitting part of the input using $\sigma$ and setting the gain $\eta(g)$ with polarization optics, the input and resource states are mixed in a polarization-based interferometer, with HWPs and BDs implementing the individual beam splitters. A small tilt in the intermediate BD introduces the necessary phase shift in the upper mode of \textbf{a}.
		The success of the protocol is heralded by a (1,1,0) photon detection pattern using high-efficiency SNSPDs.
		To validate the correct operation of the setup, the amplified output is interfered with the phase reference obtained from $\sigma$ in an imbalanced interferometer.
		\textbf{c}, By adjusting the experimental gain of the HA stage, we can compensate for the imbalance introduced by $\sigma$ and measure high-visibility interference fringes between the phase reference and the amplified output; simultaneously verifying the gain and the coherence properties of the amplifier.
		ppKTP, periodically poled potassium titanyl phosphate; BD, beam displacer; QWP, quarter-wave plate; HWP, half-wave plate; PBS, polarizing beam splitter; SNSPD, superconducting nanowire single-photon detector.
	}
	\label{fig:experiment}
\end{figure*}

\section{Results}
The conceptual idea behind HA is to coherently teleport the arbitrary state of an optical mode onto a more intense beam, resulting in a net amplitude gain. 
An $n$-photon scissor works as an ideal heralded amplifier for arbitrary quantum fields with at most $n$ photons~\cite{guanzon2022IdealQuantumTeleamplification}.
When applied to a general quantum state with the form $\ket{\psi}=\sum_{k=0}^\infty c_k \ket{k}$, written here in the photon-number basis, it truncates components beyond a finite $n$ and performs the transformation
\begin{equation}
	\label{eq:n_scissor}
	\ket{\psi}\xrightarrow{} \ket{g \psi_n} = \mathcal{N} \sum_{k=0}^{n} g^k c_k \ket{k},
\end{equation}
where $g\in[0, \infty)$ is the amplitude gain after the HA and $\mathcal{N}$ is a normalization constant.
Tele-amplification (i.e., simultaneously teleporting and amplifying a signal) happens whenever $g>1$, with success probability inversely proportional to $g^{2n}$.

The requirements for the process are a fixed ancillary resource with $n$ photons, a fixed, coherent ($n+1$)-mode interferometer, and $n$ photon detections that herald the success of the protocol (Fig.~\ref{fig:conceptual}).
The scheme is efficient in the sense that it only requires linear optics with no active components, and the required resources scale linearly with $n$. 
The gain of the amplifier is set by splitting the resource state on an optical beam splitter with transmittance $\eta \equiv \eta(g) = 1-g^2/(1+g^2)$.
The transmitted mode and the input state for the amplification are coherently combined inside a linear multi-mode interferometer.

For the $n=2$ scissor, this interferometer acts as an optical tritter~\cite{spagnolo2013ThreephotonBosonicCoalescence}, implementing a discrete, three-mode quantum Fourier transform (QFT).
Each input mode is scattered across all three output modes with a fixed phase (Fig.~\ref{fig:experiment}a).
Conditioned on the simultaneous detection of $(1,1,0)$ photons in each of the three output modes, the overall action of the 2-scissor is to transform the input state as
\begin{equation}
	\label{eq:2_scissor}
	\ket{\psi} \xrightarrow{}  \ket{g\psi_2}= \mathcal{N}\left(c_0\ket{0} +gc_1\ket{1}+g^2c_2\ket{2}\right).
\end{equation}
Here, the (unnormalized) one- and two-photon intensity gains are $G_{[1]} = |g|^2$ and $G_{[2]} = |g^2|^2$, respectively.
The combined effect of the QFT and photon detection mechanism is to subject the input state, along with part of the ancilla, to a high-order Bell measurement in the photon-number basis.
Depending on the exact detector pattern observed---(1,1,0), (1,0,1), or (0,1,1)---the input state is tele-amplified perfectly or with an additional $2\pi/3$ or $4\pi/3$ phase, respectively, that can be corrected locally~\cite{guanzon2022IdealQuantumTeleamplification}.

Our experimental setup is shown in Fig.~\ref{fig:experiment}, along with the associated circuit.
We use one type-II collinear parametric downconversion source to simultaneously generate all the photons necessary for the experiment.
With sufficiently high pump power, the downconversion process can reliably generate four photons in the polarization state $\ket{\chi} \approx \ket{2}_{H}\ket{2}_{V}$ ($H\equiv$ horizontal; $V\equiv$ vertical) in the same spatial mode.
As shown in Fig.~\ref{fig:experiment}b, we separate the two orthogonal polarizations into different modes, one of which serves as the input field to be amplified, while the other acts as the ancillary resource for the protocol.
The nominal gain $g$ of the amplifier is determined by the transmission $\eta$ of a variable beam splitter and can be freely tuned in the experiment by adjusting the polarization state of the ancilla.
Both input and resource states propagate through a large multi-mode interferometer: a passive linear optical circuit composed of half-wave plates (HWPs) and low-loss beam displacers (BDs) implementing a three-mode QFT.
Depending on the specific angle of the individual HWPs, the two modes interfere according to the circuit in Fig.~\ref{fig:experiment}a (see Methods).
At the output of the QFT, all three modes are coupled to superconducting nanowire single-photon detectors (SNSPDs), flagging the success or failure of the amplifier depending on which detectors click during the experiment.

\begin{figure*}
	\centering
	\includegraphics[width=150.34mm]{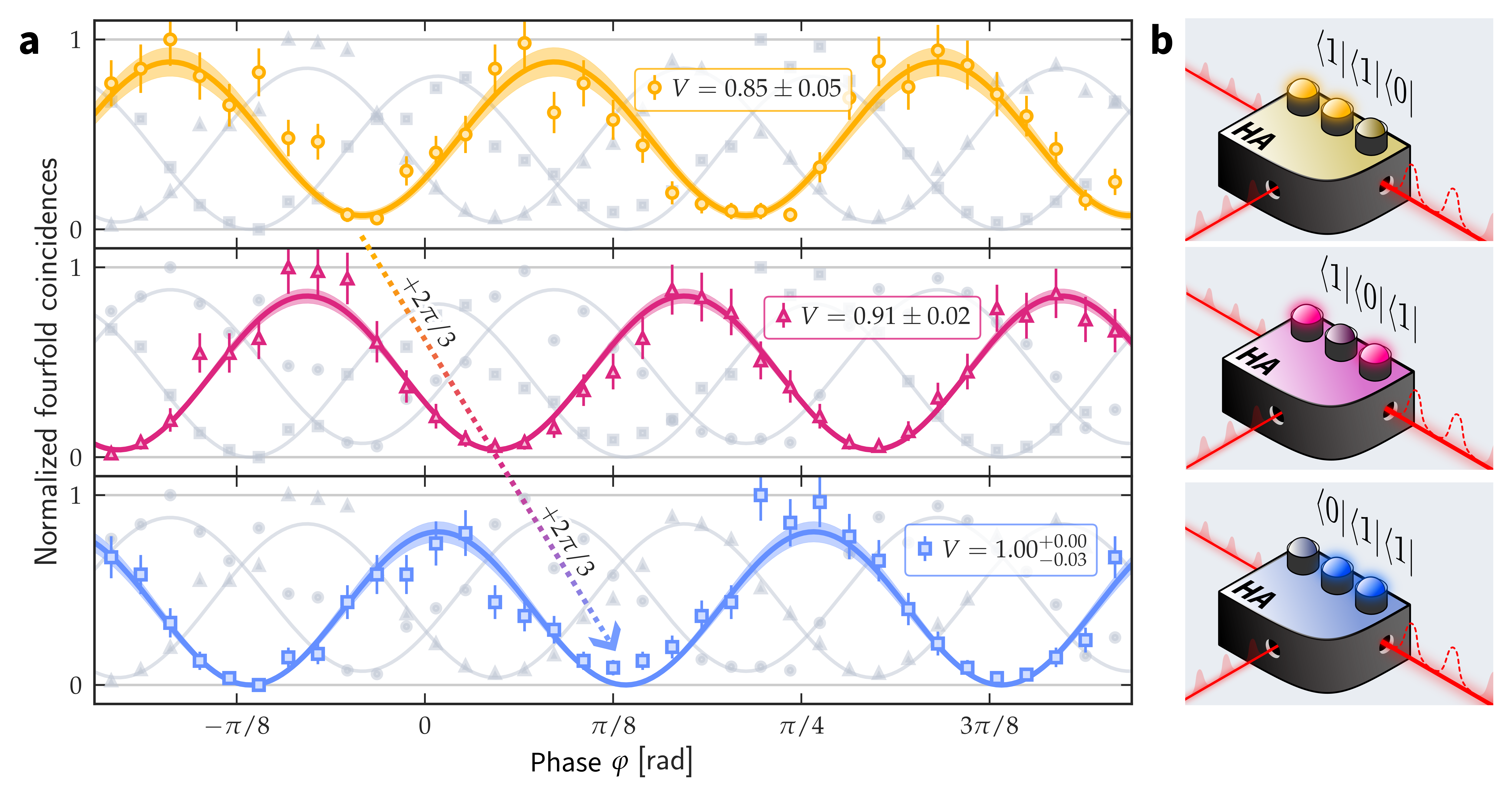}
	\centering
	\caption[]{\textbf{Coherence measurement for a nominal gain of $g=3$.}
		\textbf{a}, Heralded two-photon interference between the amplified output and a phase reference, conditioned on three possible different detection patterns.
		High-visibility fringes indicate the correct operation of the amplifier.
		\textbf{b}, Depending on which combination of detectors fire, we can observe a fixed $2\pi/3$ or $4\pi/3$ phase shift in the amplified state that can be corrected locally.
		Error bars represent $\pm 1$ s.d.~from Monte Carlo simulations under Poissonian statistics, and shaded areas correspond to $95\%$ confidence regions derived from the uncertainty in the fit parameters.
	}
	\label{fig:interference}
\end{figure*}

Ideal tele-amplification does not introduce phase noise to the output signal. 
To verify the coherence-preserving operation of the protocol, we interfere the amplified output with a well-known phase reference obtained before the amplification step.
Specifically, we prepare the state $\ket{\psi}=\ket{2}$ into an uneven superposition of paths after passing through a beam splitter with ratio $\sigma$, as per Fig.~\ref{fig:experiment}c. 
The transmitted mode is directed to our amplifier as input, while the other serves as the phase reference.
Upon receiving a successful heralding signal, we recombine the reference beam with the tele-amplified state, now in orthogonal polarizations, into the same spatial mode.
By scanning the relative phase $\varphi$ between the two polarizations, we measure the interference fringes after a 50:50 BD-splitter (see Methods for details of the coherence measurement).
In the absence of imperfections, ideal interferometric visibility requires two criteria to be met: (i) the gain must be correctly set to compensate for any initial imbalance introduced by $\sigma$, and (ii) the amplification process must be fully coherent.

We perform these measurements in an interferometric setup with a 90:10 split (Fig.~\ref{fig:interference}), requiring a nominal gain of $g=3$ (corresponding to a two-photon intensity gain of $G_{[2]} = 81$) to compensate the imbalance effects; see Supplementary Note 1 for the results for the nominal gains $g=1$ and $g=2$.
When the weak arm is successfully amplified, we measure high-quality interference visibilities of $(0.85\pm0.05)$, $(0.91\pm 0.02)$, and $(1.00^{+0.00}_{-0.03})$, determined by numerically fitting each fringe.
We also observe the fixed phase shift predicted by each of the three possible heralding patterns.
We attribute the variation in the visibilities to imperfect mode matching and experimental loss within the circuit.
High-visibility fringes further provide indirect evidence for the distillation of mode entanglement, a result that is discussed in more detail in Supplementary Note 2.

\begin{figure}
	\centering
	\includegraphics[width=88mm]{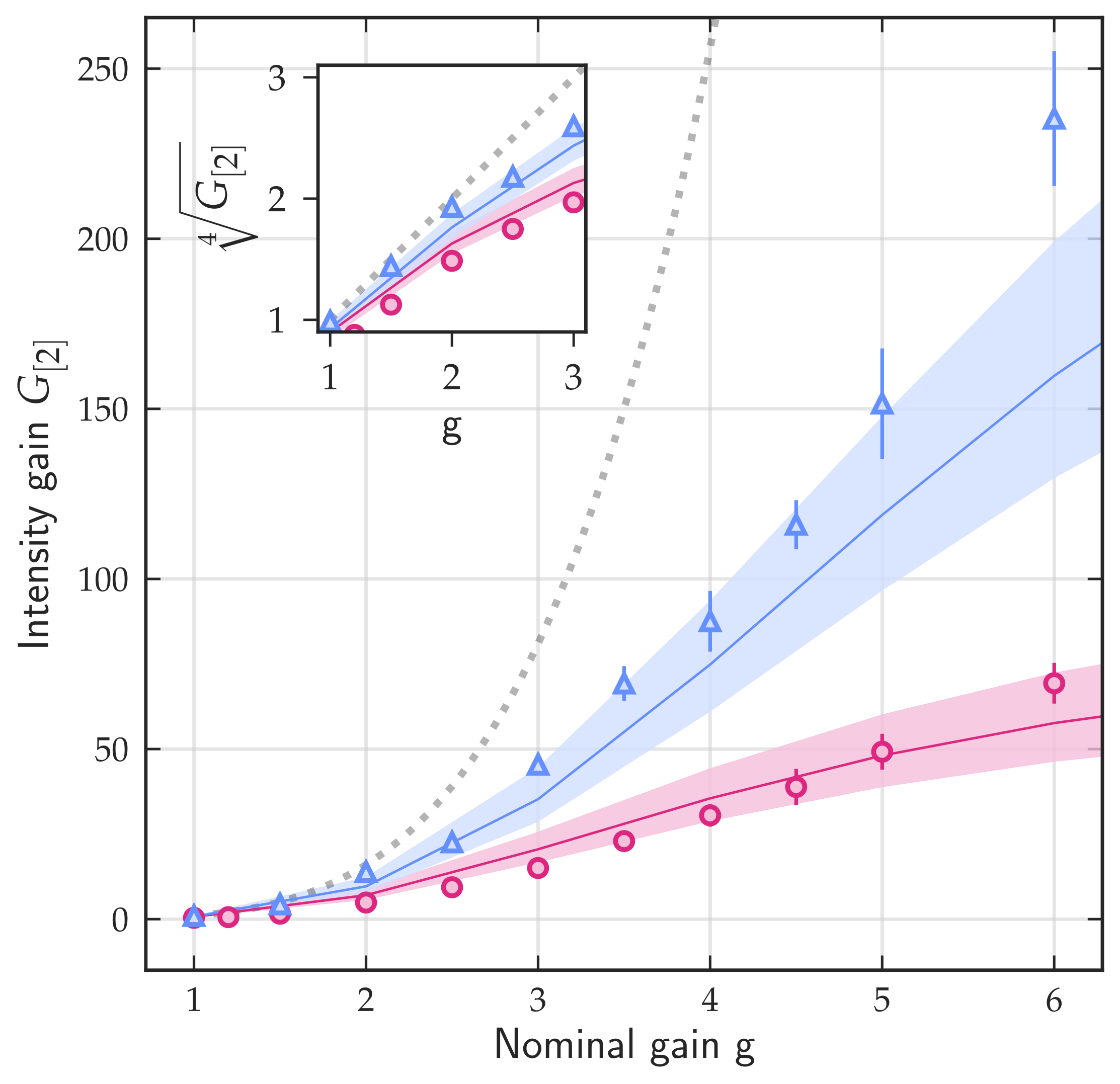}
	\centering
	\caption[]{\textbf{Measured two-photon intensity gain $G_{[2]}$ scaling under different loss channels.}
		The two-photon intensity gain $G_{[2]}$ is calculated as the ratio of the two-photon component before and after amplification, conditioned on a correct heralding signal. The nominal gain $g$ is set directly in the experiment (see text).
		Solid lines represent the theoretically modeled gains while considering small unintended losses throughout the circuit.
		Shaded regions correspond to a 95\% confidence region around theory predictions.
		As $g$ increases, $G_{[2]}$ scales quartically (dotted gray), eventually saturating due to normalizing the output state.
		Amplification after higher transmission channels ($\tau=0.1$, magenta circles) saturates earlier than lossier channels ($\tau=0.05$, blue triangles).
		Error bars represent $\pm 1$ s.d.~from Monte Carlo simulations assuming Poissonian statistics.
	}
	\label{fig:gain}
\end{figure}

An important feature of the amplifier is its ability to suppress loss-induced noise, effectively corresponding to an instance of quantum state purification.
We consider a scenario where the pure state $\ket{\psi} = \ket{2}$ is mixed with vacuum.
When applied to $\ket{\psi}$, a lossy channel with transmission amplitude $\tau$ will degrade it into a mixture of the original two-photon state $\ketbra{\psi}{\psi}$ and a noise term $\rho_\mathrm{loss}$ that consists solely of vacuum and single-photon components:

\begin{equation}\label{eq:mixed_input}
	\rho^\mathrm{in} = \left(1-\tau^2\right)\rho^\mathrm{loss} + \tau^2\ketbra{\psi}{\psi},
\end{equation}
where $\rho^\mathrm{loss} \propto (1-\tau)\ketbra{0}{0}+2\tau\ketbra{1}{1}$.

In the event of a successful heralding signal, the two-photon subspace in Eq.~$\eqref{eq:mixed_input}$ is amplified by a factor of $|g|^4$.
The noise term $\rho^\mathrm{loss}$ will transform into a slightly different state, where the single-photon component experiences a much smaller amplification of $|g|^2$ while the vacuum term remains unchanged (see Methods).
In the limit of large gain $g\xrightarrow{}\infty$, any mixed state of the form in Eq.~\eqref{eq:mixed_input} can be asymptotically purified into $\ketbra{\psi}{\psi}$.

By analyzing the photon-number distribution before and after amplification, we can directly measure the amplifier's two-photon intensity gain $G_{[2]} = \mathcal{N}|g|^4$, taking into account the normalization of the output state.
This normalization results in a saturation of $G_{[2]}$, which initially follows a quartic trend (Fig.~\ref{fig:gain}, dotted gray line) and eventually tapers off at high $g$.
As expected, saturation occurs more slowly in the high loss (low transmission $\tau$) regime.

For a loss channel with transmission $\tau = 0.05$ (Fig.~\ref{fig:gain}, blue triangles), we directly measure a two-photon intensity gain of up to $G_{[2]} = 235.3\pm16.1$, where we focus on the amplification corresponding to a (0,1,1) heralding pattern.
Additional unintended photon loss will affect the measured gain~\cite{goldberg2023TeleamplificationBorealisBosonsampling}, with photons traveling through different paths within the circuit experiencing a different overall transmission efficiency.
We theoretically model the amplifier's expected gain, considering the potential losses in the setup but without accounting for imperfect mode-matching or other sources of error (see Supplementary Note 4).
\blk

\section*{Conclusion}
Mitigating particle loss remains a crucial challenge for optics-based quantum applications. This work represents a significant advancement towards loss-tolerant approaches by experimentally extending the concept of noiseless amplification to quantum states of more than one photon in an optical mode. Our implementation of the QFT circuit is scalable to a larger number of input modes, since it relies on bulk, naturally low-loss, beam displacers that can accommodate multiple input spatial modes. With this approach, our demonstration of two-photon state amplification---the first step beyond the standard single-photon NLA---can be straightforwardly extended to the amplification of states with larger photon-number cutoffs. This includes states of different Fock components in a single mode, such as attenuated weak coherent states for long-distance quantum key distribution~\cite{korzh2015cow}, as well as highly squeezed states, which are essential in fault-tolerant continuous-variable optical quantum computing~\cite{winnel2024DeterministicPreparationOptical}. High-quality multi-photon amplification is also directly relevant for a range of quantum information tasks, such as distilling field entanglement, and is a key enabler for CV quantum repeater architectures~\cite{dias2020QuantumRepeaterContinuousvariable}.

\section*{\label{sec:methods}Methods}
\subsection*{\label{sec:spdc}
	SPDC sources}
Our single-photon source~\cite{slussarenko2017UnconditionalViolationShot} produces polarization-unentangled photon pairs via type-II spontaneous parametric downconversion (SPDC).
We use a mode-locked Ti:sapphire laser ($775$ nm central wavelength, $150$ fs pulses) to pump a 2 mm periodically poled potassium titanyl phosphate (ppKTP) crystal at 80 MHz.
Although the SPDC process yields a two-mode squeezed vacuum $\ket{\chi}\propto \sum_n\chi^n\ket{n}_H\ket{n}_V$,
the single-pair term $\ket{1}_H\ket{1}_V$ is never heralded during an experimental run since the detection scheme involves at least three photons.
Higher-order ($n>2$) terms also have a very small probability of occurring, and so we can approximate $\ket{\chi}\sim \ket{2}_H\ket{2}_V$.

To increase the probability of double-pair emission from the SPDC process, the pump power is set at $350$ mW.
To compensate for the temporal drift between polarizations in the down-conversion process, photons travel through a 1 mm KTP crystal with the optical axis rotated 90$^\circ$ with respect to the ppKTP.

\subsection*{\label{sec:qft}
	Quantum Fourier Transform}
Our QFT is a free-space, three-mode linear optical interferometer designed to implement the circuit shown in Fig.~\ref{fig:experiment}a, and is adapted from Ref.~\cite{guanzon2022IdealQuantumTeleamplification}.
The interferometer operates based on polarization, where the combined effect of HWPs and BDs functions as a tunable beam splitter.
The splitting ratios of the individual beam splitters are precisely controlled by rotating the corresponding HWPs.
To ensure the proper operation of the interferometer, we introduce an additional phase of $\varphi = 3\pi/2$ in one of the modes with a slight tilt of the second BD in the circuit.
Imperfect quantum (i.e., HOM) interference between input and resource photons within the QFT directly degrades the amplification process, leading to false positive heralding patterns that introduce unwanted noise into the output state.
To mitigate this, we calibrate the system by reducing the pump power to $\approx 50$ mW, allowing us to independently measure the HOM interference in each beam splitter of the QFT without higher-order SPDC events.
A bandpass filter with an $8$ nm full-width at half maximum is applied to remove unwanted spectral correlations between SPDC photons.
We measure the visibility of the two-photon HOM interference in the circuit to be above $0.991\pm0.001$ (see Supplementary Note 3).

\subsection*{\label{sec:phase_measurement}
	Coherence measurements}
To verify the coherence between a reference beam and the output of the HA process, we use an in-line classical interferometer, as described in the main text.
This device is conceptually equivalent to a two-path Mach-Zehnder, illustrated in Fig.~\ref{fig:experiment}c, but with orthogonal polarizations of the same spatial mode.
The interferometer works by superposing the horizontally polarized reference beam, obtained from the beam splitter after the waveplate $\sigma$, with the vertically polarized output of the amplifier.
The phase scanning mechanism consists of a rotating HWP positioned between two fixed quarter-wave plates (QWPs).
A second fixed HWP and a BD serve as a 50:50 beam splitter.
Rotating the angle of the HWP in the phase scanner by $\varphi$ modulates the relative phase between the $H$ and $V$ polarization terms of the combined optical beam.
Classical interference fringes are observed only when the $H$ and $V$ components are coherent with one another. 
When the amplitude of both beams is balanced----i.e., when the amplifier gain is appropriately set to counteract $\sigma$---the resulting fringes show perfect visibility.

For a nominal gain of $g=3$, shown in Fig.~\ref{fig:interference}, we collect a total of 2559 four-fold coincidences (using a coincidence window of $1.5$ ns) over 155 hours of continuous measurement, with minimal realignment.
To optimize interferometric visibility while minimizing loss in the circuit, we position the $8$ nm bandpass filters on the beam containing both the tele-amplified and reference modes.
Photons in the QFT are not attenuated.

\subsection*{\label{sec:gain_measurement}
	Intensity gain measurement}
To simulate a lossy channel, the input beam is attenuated before the HWP $\sigma$, creating the mixture described by Eq.~\eqref{eq:mixed_input}.
\begin{equation}
	\rho^\mathrm{in} = \left(\tau-1\right)^2 \ketbra{0}{0} + 2\tau\left(1-\tau\right)\ketbra{1}{1} + \tau^2\ketbra{2}{2},
\end{equation}

After amplification, this state is transformed as follows:
\begin{align} \label{eq:mixed_output}
	\nonumber \rho^\mathrm{out} \xrightarrow{} &\mathcal{N}\left[\left(\tau-1\right)^2 \ketbra{0}{0}\right. \\
	\nonumber &+ 2g^2\tau\left(1-\tau\right)\ketbra{1}{1} \\
	&+ \left. g^4\tau^2\ketbra{2}{2}\right],
\end{align}
where $\mathcal{N} = 1/\left[(\tau-1)^2 + 2g^2\tau(1-\tau)+g^4\tau^2\right]$.

We follow the method from Refs.~\cite{xiang2010HeraldedNoiselessLineara, kocsis2013HeraldedNoiselessAmplificationa} and use photon counting to measure the intensities of the two-photon component, $\rho_{22}$, both with and without amplification.

By taking the ratio $\rho^\mathrm{out}_{22}/\rho^\mathrm{in}_{22}$, we can obtain a direct estimate of the two-photon intensity gain $G_{[2]}=\mathcal{N}g^4$.
In the high-loss, low-gain regime, the normalization constant simplifies to $\mathcal{N}\approx 1$, and the measured two-photon intensity closely approximates $G_{[2]}\approx g^4$.
As $g$ increases, these conditions no longer hold, and a saturation effect in $G_{[2]}$ arises from $\mathcal{N}$.

Since no phase reference is required for measuring the gain, we set $\sigma = 1$ and direct the entire attenuated input signal into the circuit, where it interferes with the resource state.
The two SNSPDs previously used for coherence measurements are now repurposed as a probabilistic photon-number-resolving (PNR) detector, which is further conditioned on recording a successful heralding pattern.
When two photons arrive at the PNR detector, they probabilistically separate with 50\% probability and are detected in coincidence.
The mean output photon number $\rho^\mathrm{out}_{22}$ 
is then directly estimated from the resulting four-fold coincidences---following the normal operation of the HA with the gain determined by $\eta$---normalized by the corresponding two-fold heralding coincidences.
Although one could also consider three-fold coincidences to estimate the single-photon amplification, the probabilistic PNR detection scheme is limited by its inability to distinguish true single-photon events from instances where two photons bunch together.

To estimate $\rho^\mathrm{in}_{22}$, 
we adjust $\sigma=0$ to redirect the attenuated input to the PNR stage.
The gain HWP is set to $\eta = 1$ so that the resource state propagates through the QFT instead.
No interference between the input and resource beams occurs in this configuration.
To ensure a fair comparison, it is essential to remember that the success of the amplifier is conditioned on specific photon detection patterns, and thus $\rho^\mathrm{in}_{22}$ must be conditioned on the same heralding detections used to compute $\rho^\mathrm{out}_{22}$.

%

\section*{Data availability}
All data generated and analyzed during this study is available from the corresponding author upon reasonable request.

\subsection*{Acknowledgements}
Tahis work was supported by Australian Research Council (ARC) Grant No. CE170100012. L.V.-A.~acknowledges support by the Australian Government Research Training Program (RTP). 
\section*{Competing interests}
The authors declare no competing interests.

\clearpage
\pagebreak
\onecolumngrid
\begin{center}
	\textbf{\large Supplementary Information: A heralded quantum amplifier of multi-photon states}
\end{center}

\setcounter{equation}{0}
\setcounter{figure}{0}
\setcounter{table}{0}
\setcounter{page}{1}
\makeatletter
\def\bibsection{\subsection*{Supplementary References}} 
\renewcommand{\figurename}{{\bf Supplementary Fig.}}
\renewcommand{\tablename}{{\bf Supplementary Tab.}}
\renewcommand{\thefigure}{{\bf S\arabic{figure}}}
\renewcommand{\thetable}{{\bf S\arabic{table}}}
\renewcommand{\theequation}{S\arabic{equation}} 
\renewcommand{\@caption@fignum@sep}{\ ${\bm |}$\ }%
\renewcommand{\thesubsection}{{\bf Supplementary Section \arabic{subsection}}}
\renewcommand{\@seccntformat}[1]{\csname the#1\endcsname:\ }%
\renewcommand{\citenumfont}[1]{S#1}

\setcounter{secnumdepth}{2}

\subsection*{\label{sm-sec:fringes}Supplementary Note 1: Coherence measurements for other gain values}
\begin{figure*}[!h]
	\centering
	\includegraphics[width=112.5mm]{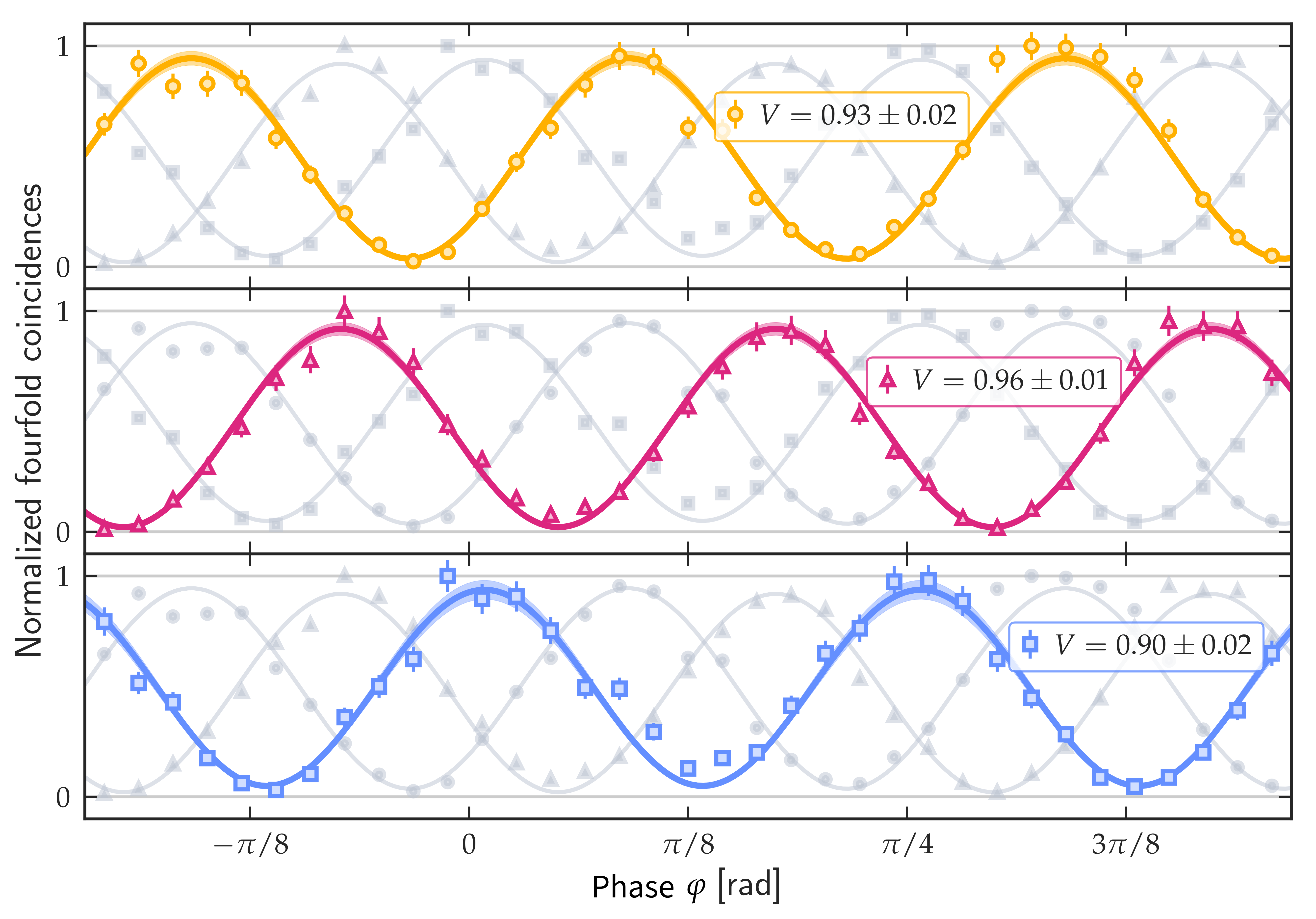}
	\centering
	\caption[]{\textbf{Coherence measurement for a nominal gain of $g=1$.}
		Heralded two-photon interference between the amplified output and a phase reference, conditioned on three possible different detection patterns.
		In this measurement, we collect 11049 four-fold coincidences over 25 hours of continuous measurement.
		The BS is set to split the input state evenly between the two paths ($\sigma=0.5$).
		Error bars represent $\pm 1$ s.d.~from Monte Carlo simulations under Poissonian statistics, and shaded areas correspond to $95\%$ confidence regions derived from the uncertainty in the fit parameters.
	}
	\label{fig:g1}
\end{figure*}

\begin{figure*}[!h]
	\centering
	\includegraphics[width=112.5mm]{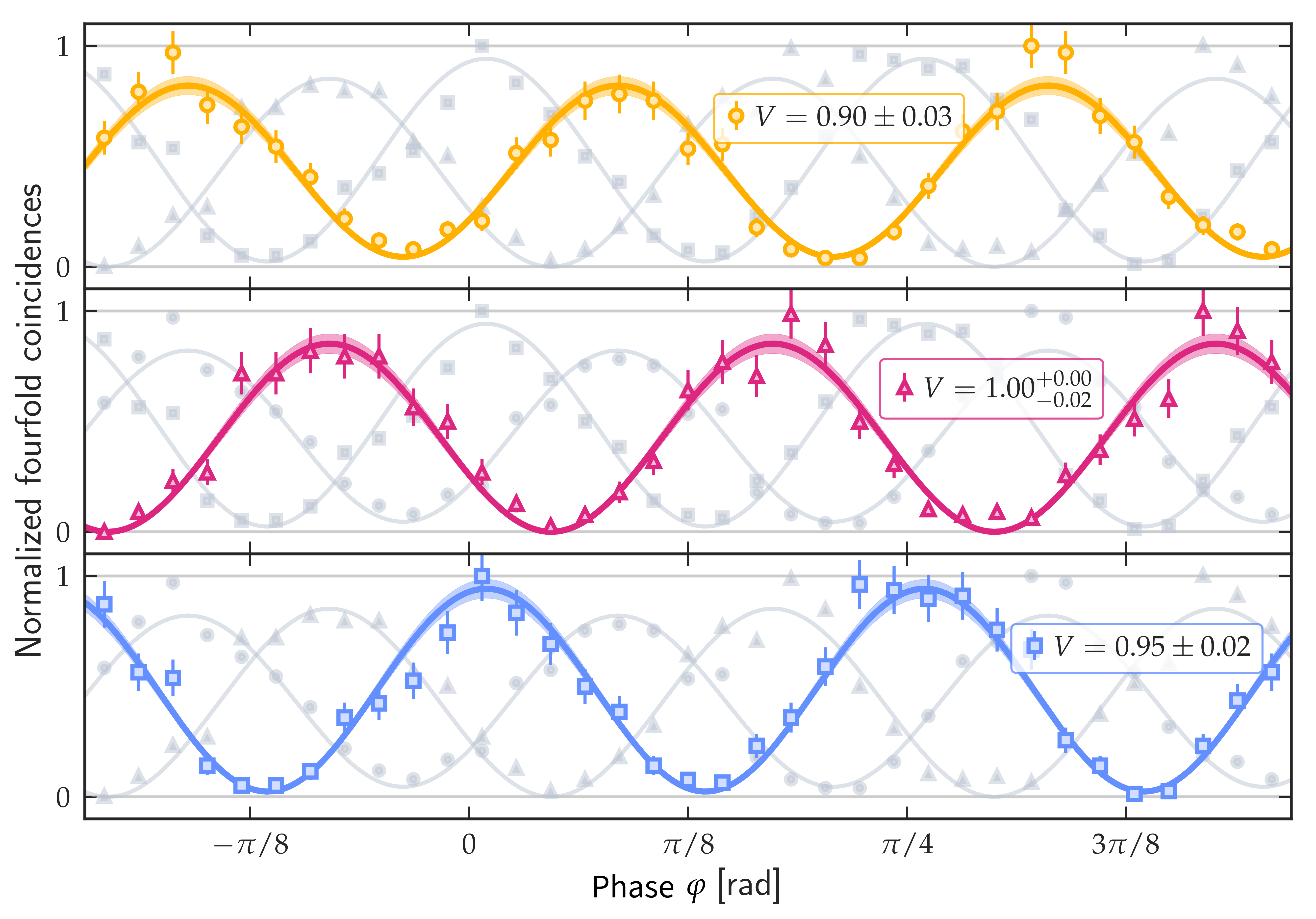}
	\centering
	\caption[]{\textbf{Coherence measurement for a nominal gain of $g=2$.}
		Same as Supp.~Fig.~\ref{fig:g1}, but for $g=2$.
		The results correspond to an 80:20 splitting ratio ($\sigma=0.2$).
		We collect 4123 four-fold coincidences over 50 hours of continuous measurement.
	}
	\label{fig:g2}
\end{figure*}

In this section, we present the recovery of high-visibility fringes for nominal amplitude gains of $g=1$ and $g=2$, using the same interferometric setup described in the main text.
The experimental setup was realigned between measurements corresponding to different gains.
We attribute the variations in fringe visibilities between different measurements to slightly different experimental conditions introduced by the realignment procedure.

\subsection*{\label{sm-sec:distillation}Supplementary Note 2: Estimation of distillable entanglement}
\begin{figure*}[!h]
	\centering
	\includegraphics[width=112.5mm]{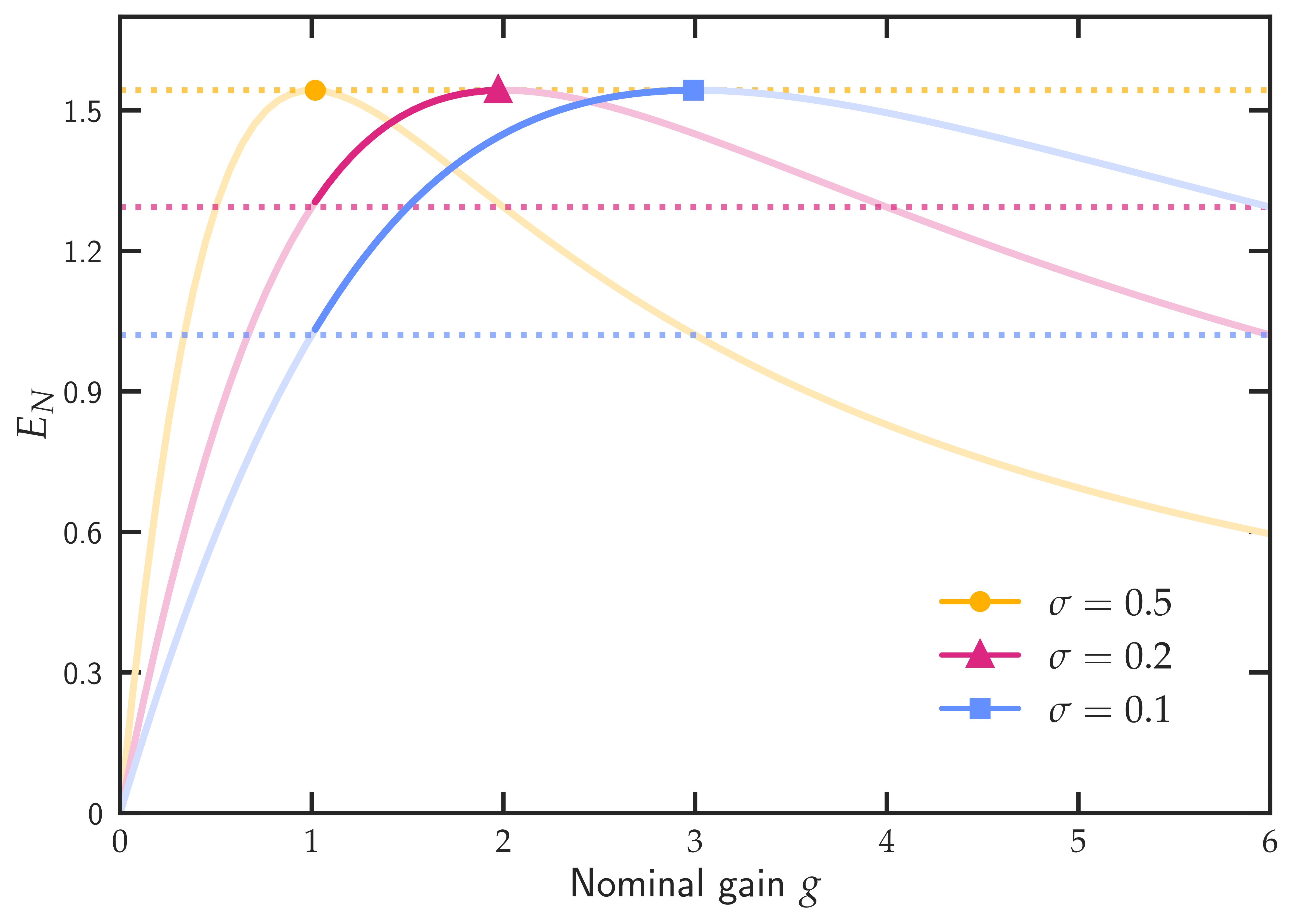}
	\centering
	\caption[]{\textbf{Theoretical analysis of the recoverable entanglement for a path-entangled $\ket{2}$ Fock state superposition.}
		The logarithmic negativity $E_N$ is plotted for both pre- (dotted lines) and post-amplification (solid curves) states, with different splitting ratios $\sigma$ as a function of the nominal gain $g$.
		Except for the case of $\sigma=0.5$ (which already has maximal entanglement), successful amplification at the HA stage leads to an increase in distillable entanglement (represented by the darker curves) up to its maximum value (markers).
	}
	\label{fig:logneg}
\end{figure*}

The results from the coherence measurement can be interpreted in the context of distilling entanglement.
Coherently splitting a $\ket{2}$ Fock state, as done in the coherence measurement using a BS with transmission $\sigma$, generates a path-entangled state in the photon-number basis:
\begin{equation}\label{eq:qutrit}
	\ket{\Psi}_{rt} = (1-\sigma)\ket{20}_{rt} + \sqrt{2}\sqrt{\sigma}\sqrt{1-\sigma}\ket{11}_{rt} + \sigma\ket{02}_{rt},
\end{equation}
where $r$ and $t$ denote the reflected and transmitted modes after the BS, respectively.
The entanglement strength depends on the splitting ratio $\sigma$, with greater imbalance leading to weaker entanglement.
Upon amplification of the transmitted arm at the HA stage, the state in Eq.~\eqref{eq:qutrit} is transformed as follows:
\begin{equation}\label{eq:qutrit_amp}
	\ket{\Psi}_{rt} \xrightarrow{} (1-\sigma)\ket{20}_{rt} + g\sqrt{2}\sqrt{\sigma}\sqrt{1-\sigma}\ket{11}_{rt} + g^2\sigma\ket{02}_{rt},
\end{equation}
up to normalization.
To estimate an upper bound on the distillable entanglement in this setup, we consider the logarithmic negativity $E_N$ for both pre- and post-amplification states under different experimental configurations.
For states generated with an imbalanced splitting ratio $\sigma$, it is always possible to compensate by increasing $g$.
This results in a direct increase in $E_N$ (Fig.~\ref{fig:logneg}, dark-colored curves), reaching its maximum for a balanced state when $g^2 = (1-\sigma)/\sigma$, denoted by the markers.
The dotted lines in Fig.~\ref{fig:logneg} represent the baseline negativity before amplification.
In the case of $\sigma=0.1$, corresponding to the experimental results shown in Fig.~\ref{fig:interference} of the main text, we theoretically estimate an increase in $E_N$ from $\approx 1.0204$ (before amplification) to $\approx 1.5431$, assuming ideal conditions without experimental imperfections.

\newpage

\subsection*{\label{sm-sec:fringes}Supplementary Note 3: HOM interference within the circuit}
\begin{figure*}[!h]
	\centering
	\includegraphics[width=112.5mm]{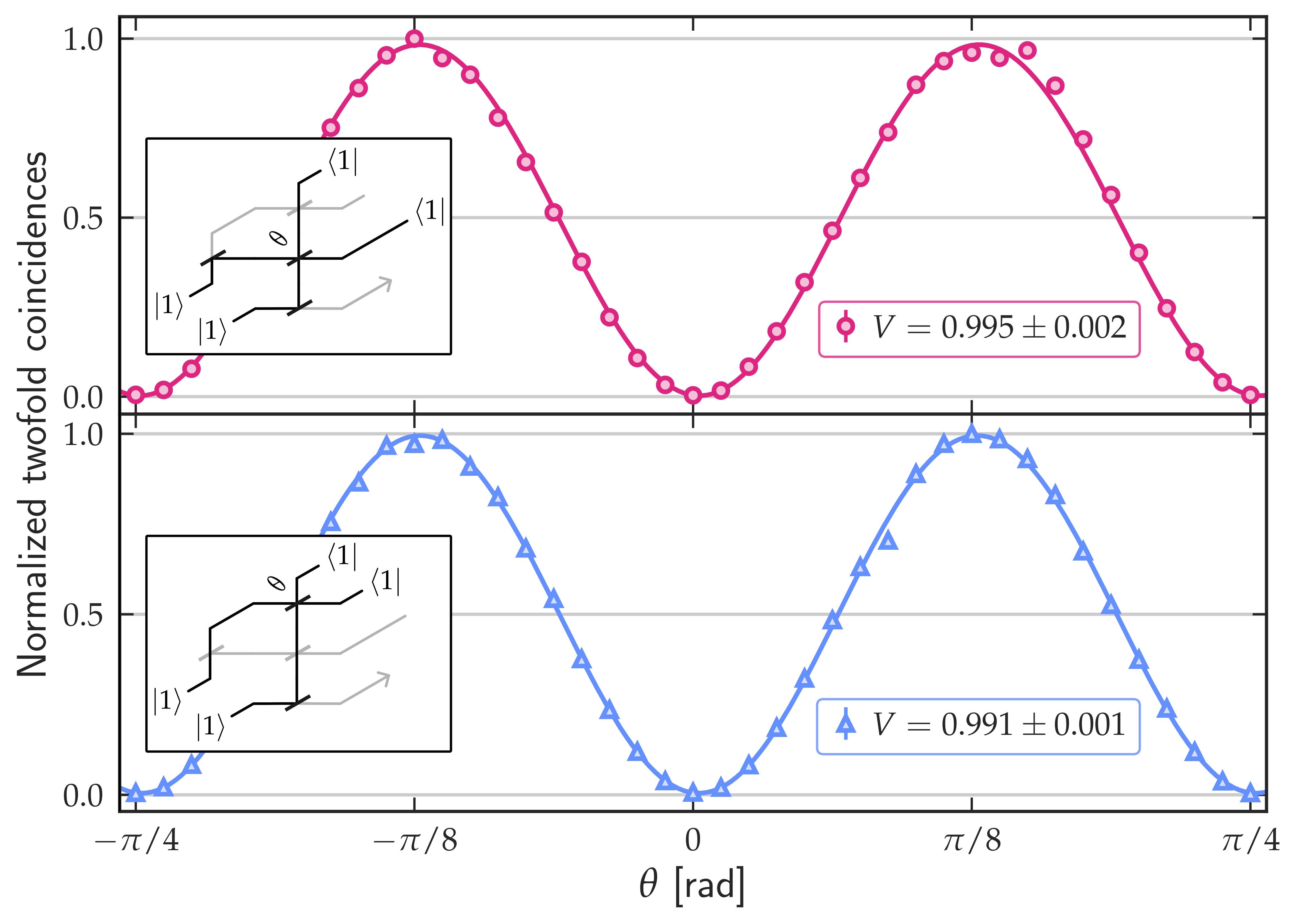}
	\centering
	\caption[]{\textbf{Hong-Ou-Mandel interference within the QFT.}
		Two-photon polarization HOM interference at two distinct locations inside the QFT (insets), as a function of the polarization angle between interfering photons.
		Error bars, smaller than the markers, represent the experimentally observed uncertainty of $\pm 1$ standard deviation, assuming Poissonian statistics.
	}
	\label{fig:hom}
\end{figure*}

Within the QFT circuit, two key points exist where the input and resource states interfere.
Given that the interferometer is polarization-based and the two states are spatially overlapped, we focus on the nonclassical HOM interference between them by rotating a HWP by $\theta$ with respect to its optical axis.
This rotation results in a two-fold coincidence probability $|\sin^2 2\theta - \cos^2 2\theta|^2$, which is plotted in Fig.~\ref{fig:hom} for the two locations (insets).
As discussed in the main text, we lower the pump beam power to $\approx 50$ mW and use mild 8 nm FWHM spectral filters.

\subsection*{\label{sm-sec:model}Supplementary Note 4: Loss model}
\newcommand{\GT}{\mathcal{G}}

\begin{figure*}[h]
	\centering
	\includegraphics[width=150mm]{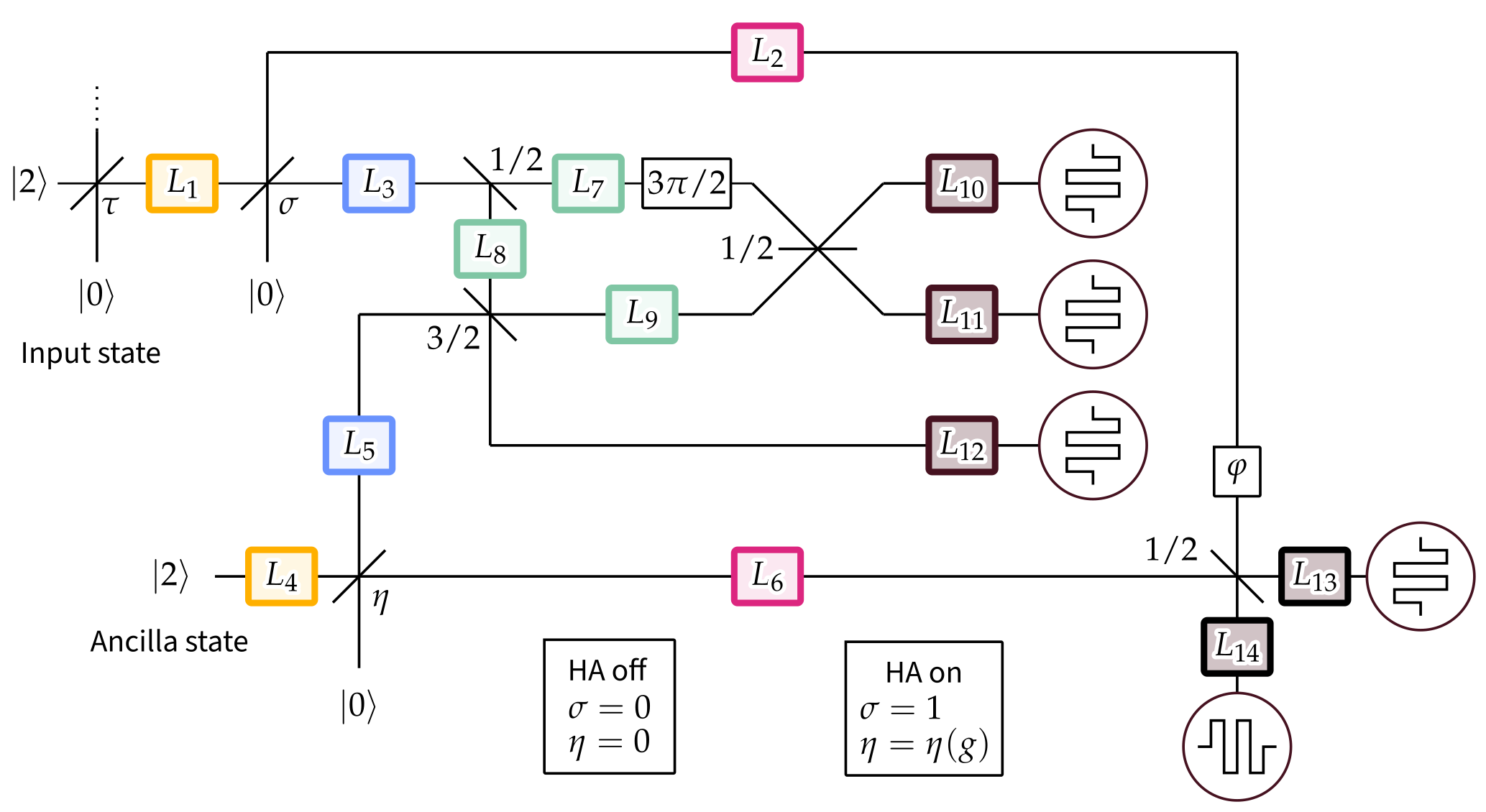}
	\centering
	\caption[]{\textbf{Theoretical loss model for the HA experiment.}
		Fourteen different loss locations $L_i$ can influence the measured two-photon gain $\GT$. Each loss location is color-coded to indicate a different role within the experiment.
		The transmission of the beam-splitters $\sigma$ and $\eta$ changes depending on whether the HA is on or off.
	}
	\label{fig:model}
\end{figure*}

As shown in Fig.~\ref{fig:model}, we account for unintended photon loss at different locations within the experimental setup.
Each loss point is represented by a variable $L_i\in[0,1]$, modeled as a beam splitter mixing the corresponding mode with vacuum.
By evolving the initial state (expressed in the Fock basis) through the circuit, we can derive an analytical estimate of the two-photon gain $G_\mathrm{[2]}$.
Let $\GT = f(g,\tau, L_1,\ldots, L_{D})$ denote the output of the model described above, which depends only on the nominal gain $g$, the channel transmission $\tau$, and $D=14$ independent loss variables.
The specific choice of $D$ is arbitrary but is physically motivated by our experimental layout.

To quantify the impact of individual losses on $\GT$, we perform a variance-based sensitivity analysis~\cite{saltelli2006SensitivityAnalysisPractices} to estimate the proportion of variance in $\GT$ that can be attributed solely to $L_i$.
In general~\cite{sobol2001GlobalSensitivityIndices}, the total variance of a black box model can be decomposed as
\begin{equation}
	\mathrm{Var}(\mathcal{G}) = \sum_i V_i + \sum_i\sum_{j>i}V_{ij} + \ldots + V_{1\ldots D},
\end{equation}
where $V_i$ represents the variance due to the individual $L_i$ and higher-order terms ($V_{ij}$, $V_{ijk}$, and so forth) denote the variance attributed to interaction between different loss terms.

The sensitivity of $\GT$ to $L_i$ is directly estimated from the so-called first-order sensitivity indices $S_i$~\cite{sobol2001GlobalSensitivityIndices}, defined as:
\begin{equation}
	\label{eq:v_i}
	S_{i} = \frac{V_i}{\mathrm{Var}(\GT)}, \quad \text{where} \quad V_i = \mathrm{Var}_{L_i}\left[\mathbb{E}_{L_{\sim i}}(\GT|L_i)\right].
\end{equation}

Here, $V_i$ is calculated from the expected value of $\GT$ when $L_i$ remains fixed but all other loss variables are allowed to vary.
Thus, $S_{i}$ physically represents the fractional contribution of $L_i$ to the observed variance in $\GT$; the closer each index is to $1$, the more sensitive $\GT$ is to photon loss at point $L_i$.
Higher-order sensitive terms can also be analyzed, though they become impractical even for moderately small $D$; for example, our simple model has $\binom{14}{2} = 91$ second-order terms.
Consequently, we concentrate only on the first-order contributions.

Since directly calculating Eq.~\eqref{eq:v_i} is analytically intractable, we statistically estimate the first-order indices from $61440$ randomly sampled loss points~\cite{saltelli2006SensitivityAnalysisPractices}.
Assuming a channel transmission of $\tau=0.05$, we repeat the sensitivity analysis for various nominal gain values.

\begin{figure*}[!h]
	\centering
	\includegraphics[width=150mm]{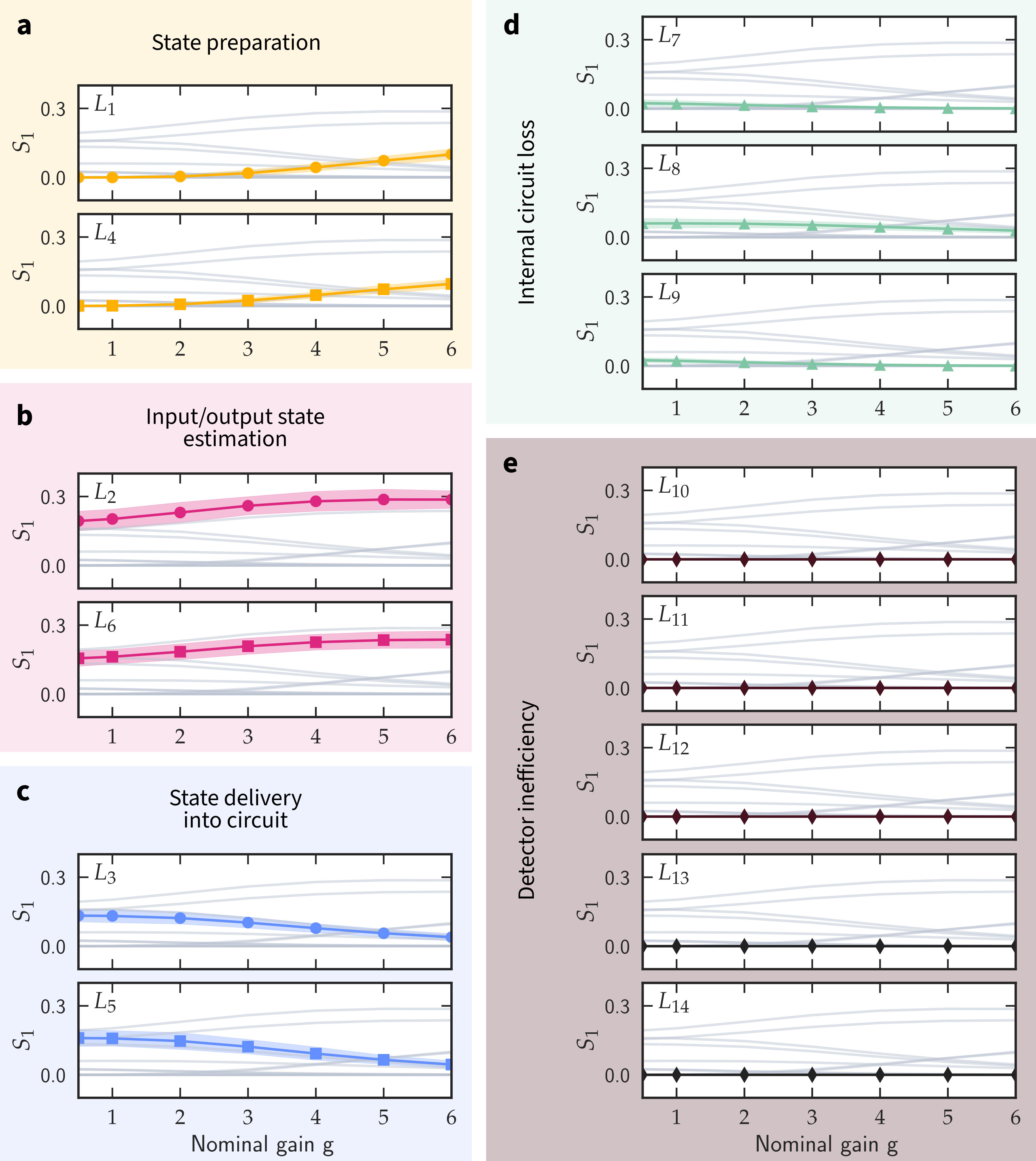}
	\centering
	\caption[]{\textbf{Global sensitivity of the amplifier to photon losses.}
		First-order Sobol sensitivity indices quantify the contribution of individual loss variables to the total variance in the measured gain, plotted as a function of the nominal gain $g$.
		Loss locations are categorized into five regions:
		\textbf{a} Post-state preparation, \textbf{b} Pre-state size measurements, \textbf{c} Pre-QFT circuit,
		\textbf{d} Within QFT circuit, and \textbf{e} Pre-detection.
		Symbols distinguish losses that affect only the input state (circles), the ancilla state (squares), or both (triangles and diamonds).
		Shaded regions show 95\% confidence intervals from the Monte Carlo analysis.
		All first-order indices are plotted for direct comparison.
	}
	\label{fig:s1}
\end{figure*}

The resulting first-order indices (Fig.~\ref{fig:s1}) quantify the influence of each loss location $L_i$ on the measured gain, with indices closer to 1 indicating stronger sensitivity.
Sensitivity analysis shows which loss locations will impact the measured gain. Still, it alone does not predict the direction of the bias, i.e., whether different losses will result in an under- or overestimation of $\GT$.
Thus, a final step is to provide some physical intuition behind the relevant loss locations.

Three distinct regions emerge as particularly significant: pre-circuit losses (Fig.~\ref{fig:s1}\textbf{a},\textbf{c}), state size estimation (Fig.~\ref{fig:s1}\textbf{b}), and losses within the QFT (Fig.~\ref{fig:s1}\textbf{d}).
Notably, detector inefficiencies (Fig.~\ref{fig:s1}\textbf{e}) have negligible sensitivity on $\GT$, with the main effect being reduced success probabilities.
The two most sensitive variables ($L_2$ and $L_6$) both physically relate to state size estimation, but with opposite effects:
Attenuating the input state ($L_2$) slightly inflates $\GT$ by making the input state appear weaker, whereas loss after amplification ($L_6$) reduces the final gain.
However, the magnitude of these losses is correlated in our experimental setup (as they occur in shared optical components), which leads to a partial cancellation of effects.
Pre-circuit losses in the input state ($L_1$ and $L_3$) effectively simulate stronger channel attenuation (i.e., smaller $\tau$).
For ancilla state losses ($L_4$ and $L_5$), we anticipate a reduction in the gain that is consistent with previous theory~\cite{xiang2010HeraldedNoiselessLineara, kocsis2013HeraldedNoiselessAmplificationa}.

\end{document}